\documentclass[emulateapj,tighten]{aastex}

\begin{document}

\title{The Spectrum of SS~433 in the $H$ and $K$ Bands}
\author{Edward L.\ Robinson\altaffilmark{1}, Cynthia S.\ Froning\altaffilmark{1}, 
        Daniel T. Jaffe\altaffilmark{1}, Kyle F.\ Kaplan\altaffilmark{1}, Hwihyun Kim\altaffilmark{1,2},
        Gregory N.\ Mace\altaffilmark{1}, Kimberly R.\ Sokal\altaffilmark{1},}
\and
\author{Jae-Joon Lee\altaffilmark{2}}
\altaffiltext{1}{Department of Astronomy, University of Texas at Austin,  C1400, 
                 Austin, TX 78712; elr@astro.as.utexas.edu}
\altaffiltext{2}{Korea Astronomy and Space Science Institute, 776 Daeduk-daero,
       Yuseong-gu, Daejeon 305-348, Republic of Korea}

\begin{abstract}
SS~433 is an X-ray binary and the source of sub-relativistic, precessing,
baryonic jets.
We present high-resolution spectrograms of SS~433 in the infrared
$H$ and $K$ bands.
The spectrum is dominated by hydrogen and helium emission lines.
The precession phase of the emission lines from the jet continues to be 
described by a constant period, $P_{jet}= 162.375$~d.
The limit on any secularly changing period
is $|\dot P| \lesssim 10^{-5}$.
The He~I~$\lambda2.0587\ \mu$m line has complex and variable 
P-Cygni absorption features produced by an inhomogeneous wind with a maximum 
outflow velocity near 900~km~s$^{-1}$.
The He~II emission lines in the spectrum also arise in this wind.
The higher members of the hydrogen Brackett lines show a 
double-peaked profile with symmetric wings extending more than
$\pm 1500$~km~s$^{-1}$ from the line center.
The lines display radial velocity variations in phase
with the radial velocity variation expected of the compact star,
and they show a distortion during disk eclipse that we interpret as a
rotational distortion.
We fit the line profiles with a model in which the emission comes
from the surface of a symmetric, Keplerian accretion disk 
around the compact object.
The  outer edge of the disk has velocities that vary from 110 to 
190~km~s$^{-1}$.
These comparatively low velocities place an important constraint
on the mass of the compact star: Its mass must be less than 
$2.2\, M_\odot$ and is probably less than $1.6\, M_\odot$.

\end{abstract}

\keywords{binaries: close --- infrared: stars --- stars: individual 
        (\objectname{SS~433}) --- stars: variables: other}

\section{Introduction}
SS~433 is an eclipsing X-ray binary star with an orbital period 
$P_{orb} = 13.08223 \pm 0.0007$~days \citep{bri05,mar13,gor11}.
One component of the binary is an A-type supergiant star 
and the other must be a neutron star or black hole accreting
mass transferred from the A-type star
\citep{bri89,gie02b,bar06,cla07,hil08,kub10}.
The system's main claim to fame is its sub-relativistic 
($V_{jet} \approx 0.26c$) baryonic jets
(for a broad review of the SS~433 jets see \citet{fab04}).
The jets originally revealed themselves as two systems
of hydrogen and helium emission lines, each system
Doppler-shifted to the red or blue by up to hundreds of
\AA ngstroms \citep{mar79a,lie79}.
Further observations showed that the Doppler shifts
of the two systems varied oppositely and symmetrically
about redshift $z \approx 0.04$ in a roughly sinusoidal
pattern
with a period near 164 days \citep{mar79b}.
\citet{mil79a} and \citet{fab79} proposed that the two systems
were produced by narrowly-columnated, oppositely-directed,
precessing jets.
A simple kinematic model based on this physical model is 
a good descriptor of the observed jet redshifts 
\citep{abe79,mar84}.
High resolution radio maps of SS~433 are in
accord with the optically-derived model \citep{rob08,bel11}.
An update to the kinematic model by \citet{eik01}
yielded precise values for the orbital inclination 
$i = 78.05^\circ \pm 0.05^\circ$, the angle between the jet and
the precession axis $\theta = 20.92^\circ \pm 0.08^\circ$, the
precession period $P_{jet} = 162.375 \pm 0.011$~days, and the 
jet speed $\beta = V_{jet}/c = 0.2647 \pm 0.0008$ or 
$\Gamma = (1-\beta^2)^{-1/2} = 1.037$.
The kinematic model only agrees with the observed jet properties
in the mean, though.
The jets ``nod'' at a period of 6.3 days,
which is the beat period between
$P_{jet}$ and $P_{orb}/2$ \citep{kat82}; 
the jets are a series of clumps or ``bullets'', not a 
continuous stream \citep{mur80,ver93};
and there is intrinsic jitter in both precession
phase and jet velocity \citep{eik01}.
The jet velocity has been attributed to line-locking in
hydrogenic ions \citep{mil79b,sha86}.
Taken together, the line locking mechanism and the strong H and 
He emission lines demonstrate that the jets are dominated
by baryons.

While the basic properties of the jets in SS~433 are well 
established, the same cannot be said about most of the
other properties of SS~433.
The existance of the jets implies the presence of a precessing
accretion disk around the compact star. 
There is abundant indirect evidence for the disk,
but direct evidence for it is meager.
The primary eclipses in the optical light curve can be understood
as eclipses of an accretion disk that precesses with, and
is perpendicular to the jets \citep{gor98,che02}.
The non-jet emission lines are generally called stationary
lines (stationary only by comparison!).
\citet{per09,per10} fitted the profiles of the 
stationary H$\alpha$, H$\beta$, and Br$\gamma$ emission line
with 4 to 6 Gaussians for each line.
Two of the Gaussians were consistently displaced by 
$\pm (500 - 600)$~km~s$^{-1}$ from the line centers.
They associated these components with the accretion disk.
\citet{fil88} successfully modeled the broad, double-peaked profiles
of the stationary emission lines in the higher members of the 
Paschen series with theoretical accretion disk line profiles, although
\citet{fab04} preferred to attribute the profiles to disk
winds and other gas outflows.

A disk wind is certainly present \citep{fab04}.
The stationary Balmer lines are dominated by emission from 
this wind, not the disk itself, because they are produced in
a volume large compared to the orbit of the
binary star and because the radial velocity variations of 
the lines are not in phase with the radial velocity 
variations expected for gas
around the compact star \citep{mur80,gie02a}.
The wind also reveals itself
in H$\beta$, He~I~$\lambda5015$, and sometimes other lines.
The wind speeds measured from the absorption components
of the P-Cygni profiles range from 
$\sim\! 150$ to $\sim\! 1300$~km~s$^{-1}$,
increasing monotonically as the disk becomes more
face on \citep{fab97a}.
The maximum wind velocities inferred from models of 
these absorption features and of the  broad wings of the 
He~II emission lines imply that the maximum wind speed
could be several times higher than $1300$~km~s$^{-1}$
\citep{fab04,med13}.
The measured mass loss rates in the wind generally 
cluster near $10^{-4}\, M_\odot\, \textrm{yr}^{-1}$
(but see \citet{kot96}).
The rate is less certain than the agreement 
among the measurements might suggest, though, because it
depends on {\it inter alia} the adopted wind speed, the
wind geometry, the clumpiness of the wind, and the mass of the 
compact star \citep{shk81,blu01,fuc02,per09}.
Nevertheless, the mass loss rate is high and connotes 
an accretion disk around the compact star that is 
supercritical and vertically extended, at least near the 
compact star \citep{fab04}.

The gas outflow is enhanced in the plane of
the disk.
The enhanced outflow is detected from absorption 
lines with radial velocities near $-100$~km~s$^{-1}$
that are strongest when the accretion disk is 
approximately edge-on \citep{cra81,fab97a,fab97b}.
This outflow is a likely source of some of the absorption
seen at X-ray wavelengths \citep{kot96}.
Since the amount of absorption depends on the jet
precession phase, \citet{kot96} invoked circumbinary 
material in a precessing plane fed by gas ``sprinkled'' 
from the accretion disk.
The geometry of the equatorial outflow is uncertain.
\citet{fab97a,fab04} pictures the material as a cometary
tail coming from the A-type star.
Or it may be an expanding spiral wave,
perhaps extruded from the outer Lagrangian point or ejected
by a propeller, or perhaps produced
by a density wake in the stellar wind \citep{saw86,wel98,kim11}.
The outflowing material in the equatorial plane is the likely
source of the equatorial ``ruff'' observed at radio wavelengths
\citep{doo09}.
In any case, there are good observational and theoretical reasons
to conclude that the circumbinary material is not confined to 
circular Keplerian orbits, and that the gas flow is affected by
more than simple gravitational dynamics.

There is no agreement on the masses of the two stars, nor
even whether the compact star is a neutron star or a black hole.
None of the properties that might uniquely identify
a neutron star such as type~I bursts or periodic flux modulations
have been observed \citep{str06,van06}, although quasi-periodic 
X-ray oscillations with a frequency near 0.1~Hz have been 
seen \citep{kot06}.
Recent measurements of the mass of the compact star range from 
$M_X = 1.45 \pm 0.20\, M_\odot$ \citep{gor11,gor13}
to $M_X \approx 16\, M_\odot$
\citep{gie02b,blu08}, and the measurements are not converging
(see \citet{kub10} and \citet{gor11} for reviews of the
mass measurements).
The higher masses would imply that the compact star is a black
hole, but the lower masses would allow it to be a neutron star.

These properties together make
SS~433 unlike any other object that has been found in the Galaxy.
While it is often included among the microquasars because of the
properties of its radio jets \citep{mir99}, the underlying system is
quite different from the other microquasars, most of which are low-mass
and intermediate-mass X-ray binaries.
Only recently has a sibling to SS~433 been found, but in another galaxy.
M81~ULS-1, an ultraluminous supersoft X-ray source in M81, has moving
emission features reminiscent of the jet lines in SS~433 \citep{liu15}.

We report here observations of the infrared spectrum of
SS~433 in the $H$ and $K$ bands at high spectral resolution,
$R= \lambda/\Delta \lambda \approx 45,000$,
and high signal-to-noise ratio (S/N).
The previous literature on the infrared spectrum of SS~433 is sparse.
\citet{tom79} observed SS~433 in the $H$ and $K$ bands at a resolution
near $R \approx 2500$ and with a maximum S/N of $\sim\! 10$.
They detected the P$\alpha$ jet line and stationary emission
lines of He~I and the hydrogen Brackett series.
\citet{mca80} and \citet{fuc02} obtained spectrophotometry 
of SS~433 from 2.0 to 12 $\mu$m with good S/N but 
$R \lesssim 100$.
\citet{per09} observed SS~433 on 14 nights, mostly in the 
$K$ band at a resolution of $\sim\!3900$ and with
S/N $\approx 400$ at Br$\gamma$, but published only
the Br$\gamma$ data.
Section 2 of our paper describes the observations and
Section 3 gives the line identifications for both the
jet and the stationary lines.
Section 4 discusses the disk model for the higher members
of the Brackett lines.
Section 5 summarizes and discusses our results.

\section{Observations and Data Reduction} \label{sec_observ}

The Immersion GRating INfrared Spectrograph (IGRINS)
measures infrared spectra simultaneously in the $H$- and $K$-bands
(1.46-1.81~$\mu$m and 1.93-2.47~$\mu$m) 
at a resolving power of R$\simeq$45,000 \citep{park2014}.
We observed SS~433 five times between 2014 May and 2015 July with IGRINS 
on  the 2.7-m Harlan J. Smith Telescope at McDonald Observatory.  
The times of the observations are listed in Table~\ref{Obs-tab}. 
Most observations were obtained in an ``ABBA" observing sequence with 300~s exposure 
times for each individual exposure giving a total of 20 minutes on target. 
The exceptions are the 2014 May 24 observations, which were obtained in 
two ``AB'' sequences of 480~s each for a total of 32 minutes.
A nearby A0 star was observed regularly to provide telluric correction, and
calibrations (dark, flat field, and Th-Ar arc lamp exposures) were 
obtained nightly.

The data were reduced using a dedicated Python-based IGRINS data reduction 
package \citep{lee15}.\footnote{see also https://github.com/igrins/plp.}
The pipeline performs image calibration, spectrogram extraction, and wavelength 
calibration. 
We edited the comparison star spectrograms to flatten the intrinsic absorption 
features of the A star, and
then divided the SS~433 spectrograms by the A star spectrograms to remove telluric 
absorption and correct for the blaze profile of the spectral orders.
We eliminated wavelengths where the signal-to-noise
ratio of the A star spectrogram was $<20$ and/or the signal
in the normalized SS~433 spectrogram dropped below 0.05. 
Finally, we rebinned the spectrograms to a linear dispersion (0.00003~$\mu$m/pixel or 
roughly 1 resolution element) and combined the orders using weighted averages.

There are two primary periodicities in SS~433: the orbital period
and the jet precession period.
We calculated the orbital phases at which our data were obtained using
the orbital ephemeris determined by \citet{gor11} 
from data obtained between 1978 and 2007.
In this ephemeris $\phi_{orb} = 0$ corresponds to the time of
minimum of the deeper of the two optical eclipses.
The orbital phases are listed in Table~\ref{Obs-tab}.
When extrapolated forward to the epochs of our observations, the
formal error on $\phi_{orb}$ is small, $\pm0.004$, but the 
eclipsed object is a shape-shifting and aspect-shifting accretion disk, so 
observed times of minima scatter by more than 
$\pm0.2$~d ($\Delta \phi_{orb} = \pm0.015$) about the calculated times.
More significantly, as shown in Figure~4 in \citet{gor11}, the mean times of 
eclipse in the RXTE ASM data occur systematically $\sim\!0.027$ in phase
after the mean times of eclipse in the optical data.
If the X-ray emission arises close to the compact star,
this suggests that true orbital conjunction of the stars occurs at orbital phase 
$\Delta\phi_{orb} \approx 0.027\pm0.004$ after the time of eclipse
given by the Goranskij ephemeris.
This seemingly small difference between the times of true conjunction and
eclipse will become important when
we discuss the profiles of the Brackett lines.

To calculate the expected phases and redshifts of the jet lines we 
used the jet ephemeris from \citet{fab04}.
This is the same as the jet ephemeris determined by \citet{eik01}
except that phase zero is shifted to the times
when the jet points most directly at the Earth (when the H$\alpha-$
line reaches maximum blueshift), which are also the
times when the positive and negative jet lines are most separated
in wavelength.
We denoted the phase calculated from the \citet{fab04} version
of the ephemeris by $\Psi_3$ (see Figure~4 in \citet{fab04}
and Figure~1 in \citet{gor11}).
The jet phases at the times of our observations are also listed
in Table~\ref{Obs-tab}.

\section{Line Identifications}

\subsection{The Stationary Lines}

\begin{figure}
   \figurenum{1}
   \includegraphics[angle=-90.0,scale=0.65]{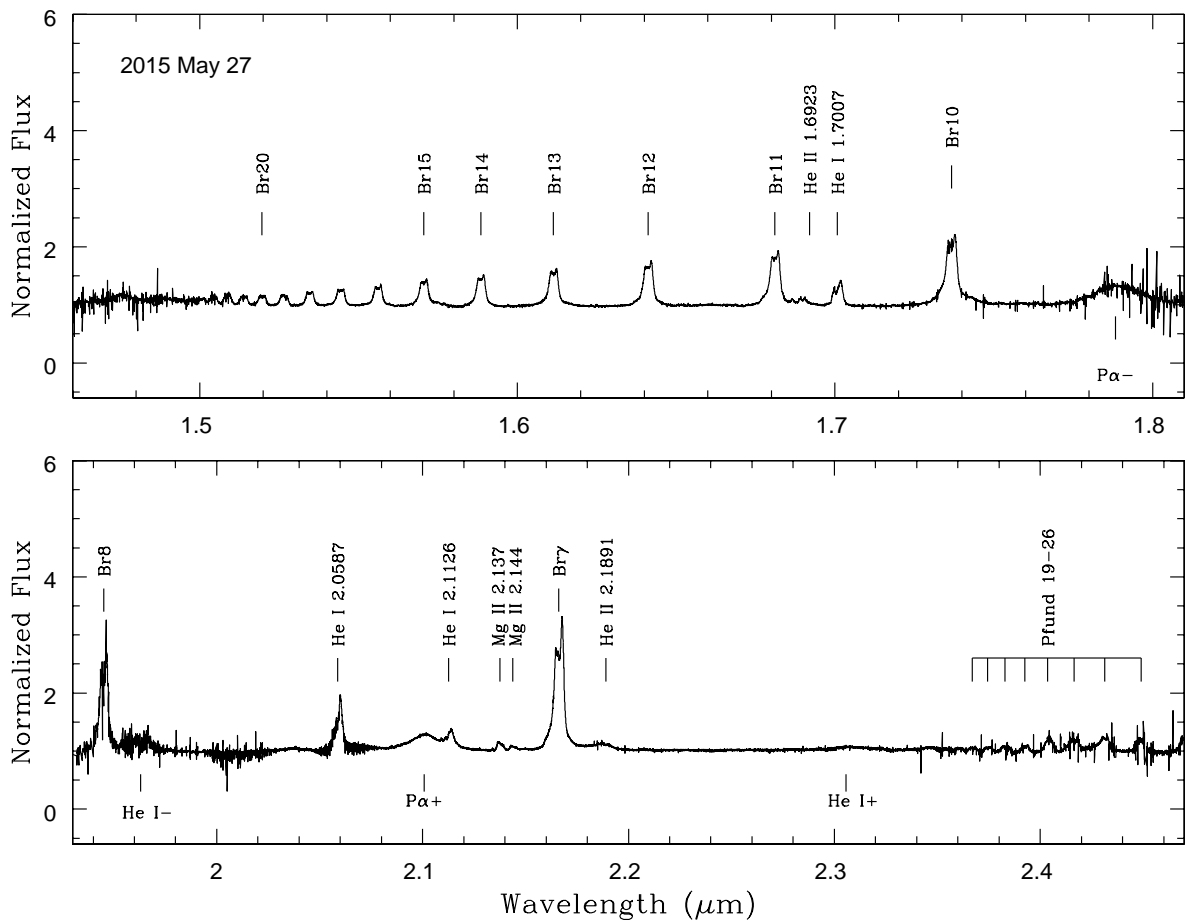}
   \caption{The spectrum of SS 433 obtained on 27 May 2015.  
    The jet lines were unusually weak at that time, allowing the stationary 
    lines to be identified without confusion.  
    The stationary lines are marked and identified above the spectrum.
    The jet lines are marked and identified below the spectrum.}
    \label{May27Spectrum-fig}
\end{figure}

Figure~\ref{May27Spectrum-fig} shows the infrared spectrum of SS~433
on 2015~May~27 when the jet lines were weak, allowing the stationary 
emission features to be identified without confusion.
The spectrum shows lines in the hydrogen Brackett series and Pfund series, 
neutral and ionized helium, and the magnesium
doublet at $\lambda\lambda2.137-2.144\ \mu$m (see Table~\ref{LineIDs-tab}).
Fe~II lines are often observed in the infrared spectra of early-type
emission-lines stars, but are absent or too weak to be detected in
the spectrum of SS~433 \citep{cho15}.
Nor do we detect the Na~I doublet at
$\lambda\lambda2.206$-$2.209\ \mu$m or any forbidden lines.

\begin{figure}
   \figurenum{2}
   \hspace*{1.4in}
   \includegraphics[scale=0.50]{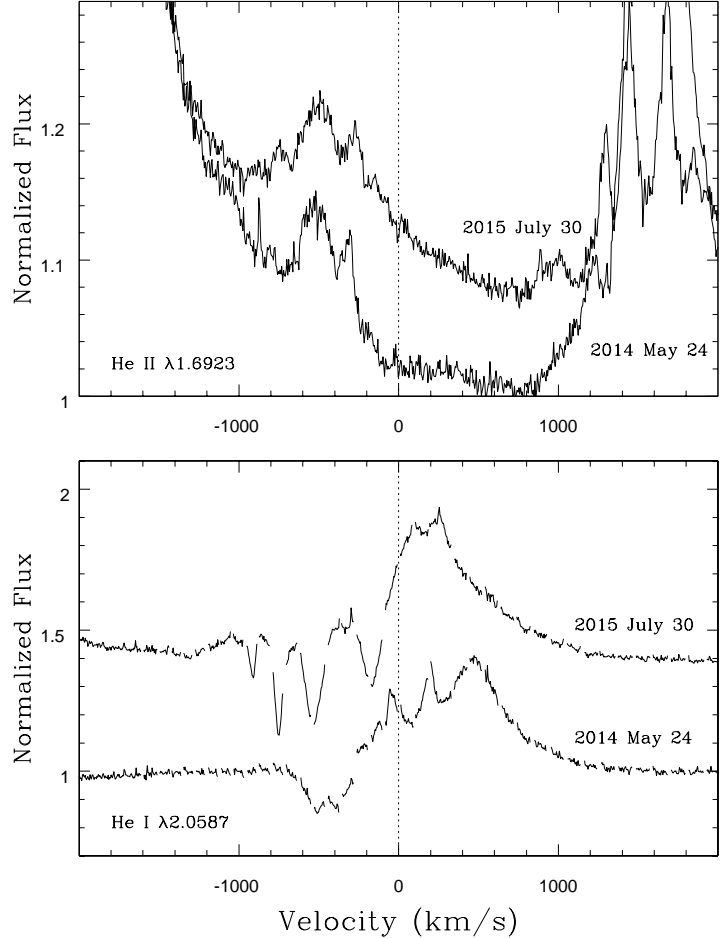}
   \caption{The bottom panel shows the profiles of the He\ I $\lambda2.0587~\mu$m
      line on 2015~July 30 when the accretion disk was nearly edge on 
      ($i = 98^\circ$), and
      2014~May~24 when the disk was most nearly face on ($i = 63^\circ$). 
      The telluric lines have been masked out in both spectrograms and the
      spectrum on 2015~July~30 has been shifted vertically for clarity.
      The wavelength scale has been converted to velocity shift from the 
      rest wavelength of the line.
      The blue-shifted absorption features betray the presence of a complex
      and variable wind outflow.
      The top panel shows the profiles of the He\ II $\lambda1.6923~\mu$m 
      line on the same dates and, once again, the spectrum on 2015~July~30 
      has been shifted for clarity.
      This weak line lies on the steeply sloping red wing of the strong Br11 
      line.
      The He~II line is shifted and broadened to roughly the same velocities 
      as the wind features in the He~I line, showing that it, too,  arises
      in the disk wind.}
   \label{Windlines-fig}
\end{figure}

The He~I~$\lambda2.0587\ \mu$m emission line, which arises from the
metastable 2s$^1$S state, shows variable P-Cygni absorption features,
the unambiguous signature of wind outflow.
The lower panel of Figure~\ref{Windlines-fig} shows the He~I line on 
2014~May~24 and 2015~July~30.
The horizontal axis in the figure has been converted to velocity shift
with respect to the line's rest wavelength.
The P-Cygni absorption features are variable and complex, showing
that the wind is highly inhomogeneous.
Individual components of the wind outflow can have
velocities up to $\sim900$~km~s$^{-1}$.
These results agree qualitatively with earlier measurements
based on absorption lines.
\citet{fab97a}, for example, deduced wind speeds between 200 and 
1300~km~s$^{-1}$ (see Figure~18 in \citet{fab04}).
In addition, Figure~5 in \citet{gie02a} suggests wind speeds between $\sim200$ and 
$\sim 600$~km~s$^{-1}$ from the P-Cygni absorption in the
He~I~$\lambda$5876~\AA\ line.
We do not, though, see the strong dependence of wind speed and
absorption line strength on the disk inclination reported
by \citet{fab97a}.
The disk was nearly edge-on on 2015~July~30 and nearly face-on
on 2014 May 24 (see Table~\ref{ModelParams-tab}), but the line 
strengths differed by only a factor of two and the maximum wind speeds
differed by just $\sim 200$~km~s$^{-1}$.

The emission feature near $\lambda1.6898\ \mu$m was an interesting
puzzle.
An emission feature at a similar wavelength is present in the spectra
of many early-type emission-line stars and is usually identified as 
C~I~$\lambda1.6895\ \mu$m \citep{cho15}.
\citet{med13} and \citet{gor11} have shown that the wind should and
does produce He~II emission lines.  
Prompted by their work, we instead identify the feature in 
SS~433 as Doppler-shifted
He~II~$\lambda1.6923~\mu$m arising in the high-velocity wind.
To justify this identification, we note that
\begin{itemize}
\item The C~I line in the spectra of Be stars is almost always weaker than
Fe~II~$\lambda1.6878\ \mu$m, usually much weaker
(see, eg, Figure~5 in \citet{cho15}).  
Neither that Fe~II line nor, for that matter, any other Fe~II lines are visible in
the IGRINS spectrograms of SS~433.
Thus, if identified as C~I, the feature at $\lambda1.6898\ \mu$m would 
be anomalously strong relative to other metal lines.

\item Although emission lines from carbon are seen in the
optical spectrum of SS~433, they are always lines of ionized
carbon(eg, C~II $\lambda7231/7236$~\AA\ and the
C~III blend near $\lambda4650$~\AA), not neutral carbon
\citep{lie79,gie02a,gie02b,sch06,med13}.

\item If identified as He~II, the feature 
has the wavelength and profile expected for emission from the wind.
The top panel of Figure~\ref{Windlines-fig} shows the profiles of 
the feature in the same spectrograms as the He~I lines
in the bottom panel of the figure.
The horizontal axis in the top panel is again the velocity shift
with respect to the rest wavelength of the He~II~$\lambda1.6923~\mu$m line.
The $\lambda1.6898\ \mu$m emission feature lies on the steeply-sloping
red wing of the strong Br11 line so its profile is distorted.
Emission lines from the wind come from a different 
volume of the wind than that which causes the P-Cygni absorption.
Recognizing that the  profiles of emission and absorption lines from an
inhomogeneous wind will not, therefore, mimic each other precisely,
we see that the $\lambda1.6898\ \mu$m emission feature 
aligns well with the P-Cygni absorption feature.

\item The $\lambda1.6923\ \mu$m line is produced by the 12-7 transition of He~II.
The He~II at $\lambda2.1891\ \mu$m is produced by the 10-7 transition.
If our identification is correct, the $\lambda2.1891\ \mu$m line must also be
present and should be stronger than the $\lambda1.6923\ \mu$m line.
Reference to Figure~\ref{May27Spectrum-fig} shows that the $\lambda2.1891\ \mu$m line
transition is, indeed, present and that it, too, is distorted and 
shifted, although not by as much as the  $\lambda1.6923\ \mu$m line.
\end{itemize}

Our identification of the $\lambda1.6898\ \mu$m has important implications.
Since the red wing of the line is weaker than the blue wing, something,
presumably the accretion disk, is obscuring the wind outflow at 
positive velocities.
Also, since the He~II emission comes from an inhomogeneous, variable,
partially obscured wind, it is dangerous to use the velocity of the He~II
lines as a proxy for the orbital radial velocity variations of the compact star.

\begin{figure}
   \figurenum{3}
   \hspace*{1.4in}
   \includegraphics[scale=0.50]{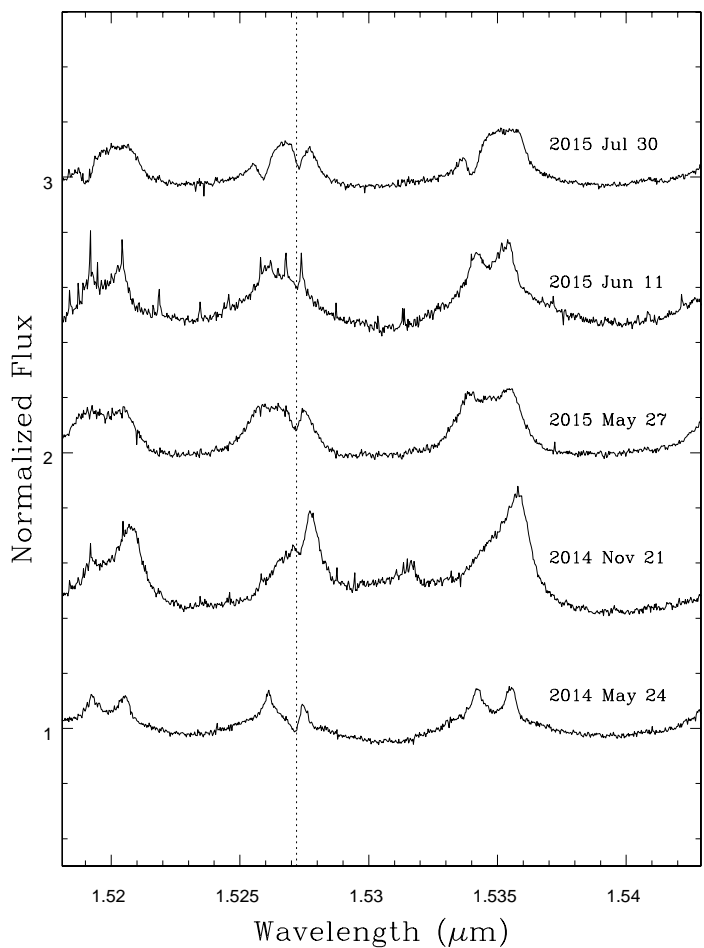}
   \caption{The Br18 (rightmost), Br19, and Br20 emission lines in all five 
      spectrograms, labeled with the dates on which the data were obtained.
      The forest of narrow features present in all the spectrograms, but 
      especially evident in the spectrogram obtained on 2015~June~11, are
      caused by incompletely removed terrestrial absorption lines.
      The profiles of all the Brackett lines are generally similar, as can be
      seen by comparing the profiles of the Br18 and Br20 lines.
      The broad absorption feature near 1.5273~$\mu$m that distorts 
      the profile of Br19 is a diffuse interstellar band.}
   \label{DIB-fig}
\end{figure}

The only absorption features other than the P-Cygni absorption that 
we could identify with certainty are two diffuse interstellar absorption 
bands (DIBs) at 1.5273 and $1.5673\ \mu$m \citep{geb11}.
The DIB at $1.5273\ \mu$m is shown in Figure~\ref{DIB-fig}, where it
shows up as an absorption feature in the Br19 line that is not
present in the Br18 or Br20 lines.

\subsection{The Jet Lines}
\begin{figure}
   \figurenum{4}
   \includegraphics[angle=-90.0,scale=0.65]{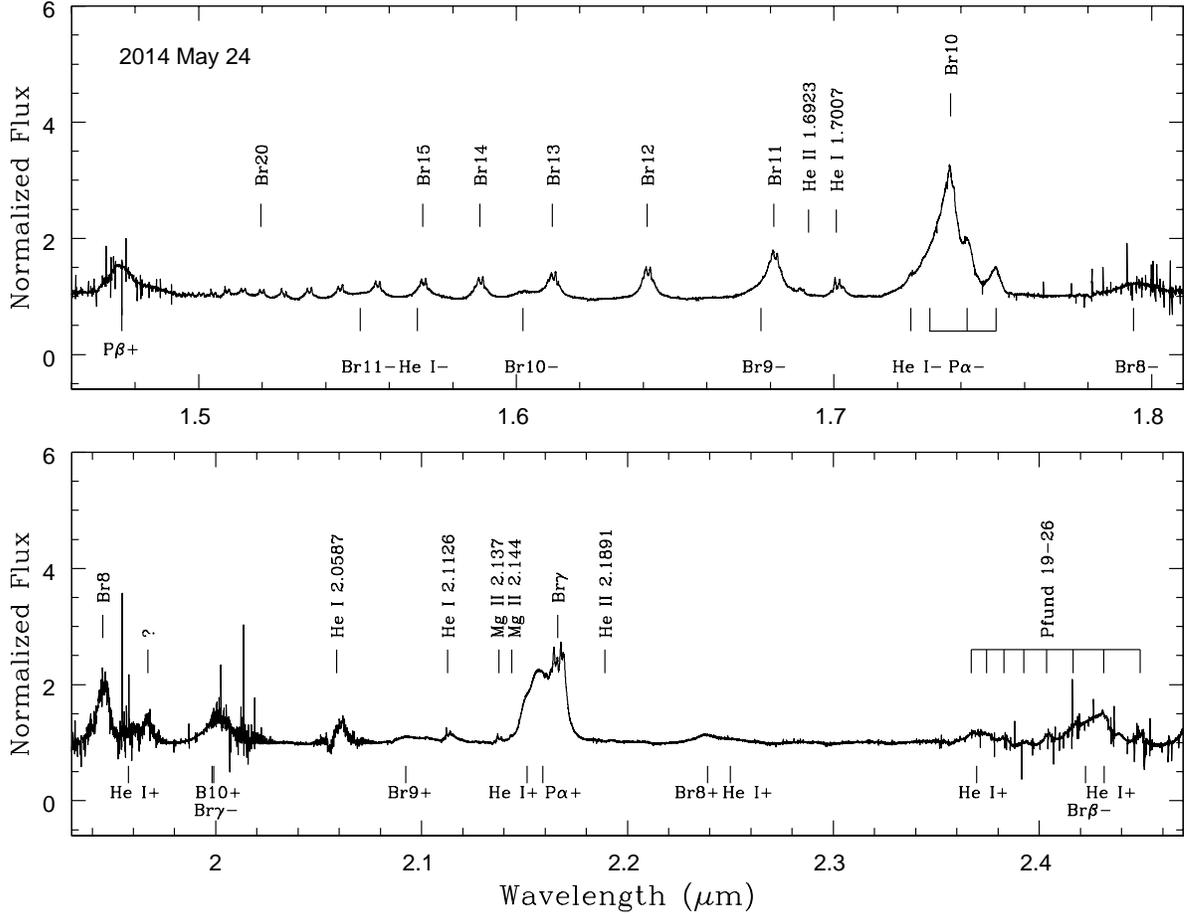}
   \caption{The spectrum of SS 433 obtained on 24 May 2014.
      The stationary lines are marked and identified above the spectrum,
      while the jet lines are marked and identified below the spectrum.
      The velocity shifts of the jet lines placed 
      the P$\alpha-$ and P$\alpha+$ lines on top of the stationary
      Br10 and Br$\gamma$ lines respectively, greatly distorting
      the profiles of those stationary lines.
      We interpret the multi-peaked profile of the P$\alpha-$
      line sitting on top of Br10 as caused by
      ``bullets'' in the jet.}
   \label{May24Spectrum-fig}
\end{figure}

Figure~\ref{May24Spectrum-fig} shows the infrared spectrum of SS~433
on 2014~May~24 when the jet lines were strong.
All the jet lines we identified in this and the other four spectrograms
are listed in Table~\ref{LineIDs-tab}, and come from neutral hydrogen and helium.
The list includes lines that are not normally in the IGRINS bandpasses
but can move into the bandpasses when the jet Doppler shifts are large.
We see one likely case of bullets in the jets:  The P$\alpha-$ line on 2014~May~24 
(Figure~\ref{May24Spectrum-fig}) has three peaks that we interpret as
coming from three different jet bullets.
The half-width at half maximum of the P$\alpha$ jet line falls in the
range 1000-2000\ km\ s$^{-1}$, varying considerably from observation to 
observation.
Although wide in an absolute sense, the jet lines are narrow compared to the
total range of projected jet velocities over the 162-day precession period.
According to \citet{ver93}, the jet bullets are typically separated by
intervals of $<1$~day, so the jet lines usually include emission from more than
one bullet.
As noted by \citet{ver93}, the narrowness of the jet lines then means
the individual bullets in the jet must cease producing
line emission within a day or so after being ejected, traveling
just $\sim10^{15}$~cm or a few hundred AU before fading.

\begin{figure}
   \figurenum{5}
   \hspace*{0.75in}
   \includegraphics[angle=-90.0,scale=0.50]{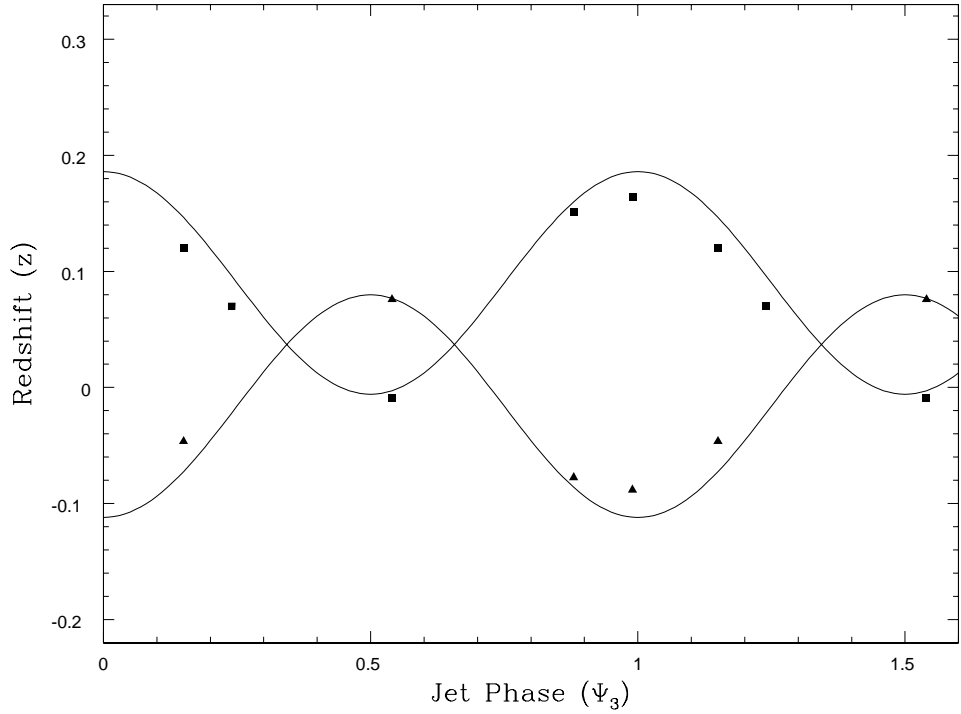}
   \caption{The filled squares and triangles are the measured
            redshifts of the jet lines in our data.
            At the scale of this plot the error bars are smaller
            than the symbols.
            The curves are the redshifts predicted by the
            \citet{eik01} model for the jet reshifts.
            A shift of the predicted redshift curves by 
            $\Delta \Psi_3 \approx -0.03$ 
            or about 5 days
            would bring the observed and predicted redshifts
            to the same phase.}
   \label{JetRedShifts-fig}
\end{figure}

The mean observed redshifts of the jet lines are given in 
Table~\ref{RedShifts-tab} along with the the values of $\Psi_3$ calculated
from the \citet{fab04} ephemeris.
The observed redshifts correspond to the peaks of the jet lines.
The peaks are often difficult to identify for individual
lines, as in the case
of the P$\alpha\pm$ lines in Figure~\ref{May24Spectrum-fig}, but
the mean redshifts are accurate to better than $\pm0.001$.
Figure~\ref{JetRedShifts-fig} is a plot of the observed redshifts along with 
the redshift curves predicted by this ephemeris.
Shifting the predicted redshift curves by 
$\Delta \Psi_3 \approx -0.03$ or about 5 days would be enough to bring them
into agreement with the observed redshifts.
Since jet nodding and intrinsic phase jitter can introduce phase
residuals of 5 to 10 days \citep{eik01}, and since we have only five
epochs, this small shift could have been introduced by statistical
fluctuations.
The most we can conclude is that the \citet{eik01}/\citet{fab04} ephemeris
predicts of the phase of the jet redshifts to
$|\Delta \Psi_3| \lesssim 0.03$.

The line identifications listed in Table~\ref{LineIDs-tab} account
for all the strong emission features in the IGRINS spectrograms of 
SS~433 except for a feature near $1.966\ \mu$m that appeared 
on 2014~May~24 and 2015~July~30 but not in our other spectrograms
(marked with a ``?'' in Figure~\ref{May24Spectrum-fig}).
Although in a region with much telluric absorption, the feature is
strong and likely to be real.

\section{A Disk Model for the Stationary Hydrogen Brackett Lines}

\subsection{Previous Models for the Hydrogen Line Profiles}

Most discussions of the stationary hydrogen emission lines have relied 
almost exclusively on observations of H$\alpha$.
The line is formed primarily in the disk wind and the wind emission 
is so strong that it tends to drown out emission from other sources.
There are occasional double peaks near the center of H$\alpha$ that
\citet{blu08} and \citet{bow11,bow13} attributed to a rotating disk.
Because they did not detect orbital radial velocity variations
in the double peaks,
\citet{blu08} concluded that the disk is a circumbinary disk or ring.
\citet{bow10} suggested that H$\alpha$ emission from an inner accretion disk 
around the compact object became visible only during a flare in 
early 2004 November.

The Paschen and Brackett lines appear to be less dominated by the wind emission,
allowing emission from other parts of the SS~433 system to be more 
visible.
\citet{per09} fitted Br$\gamma$ with six Gaussian distributions, which they 
attributed variously to a multicompenent wind, an inner accretion disk,
and an outer accretion disk.
The model fits the line profile well, but its functional form is not
based on a physical model
and it has at least 19 parameters (three for each Gaussian and one for the 
continuum), so its uniqueness is an issue.
\citet{fil88} observed the higher members of the Paschen
series.
The lines varied rapidly but on average were double-peaked with
broad wings.
\citet{fil88} modeled the mean line profiles with 
emission from a single, Keplerian accretion disk around the 
compact star.
While the fits were not perfect, the model is physically based,
testable, and requires only 6 parameters.

\subsection{A Disk Model for the Brackett Lines}

Like the Paschen lines, the Brackett lines are usually double-peaked
and have broad wings.
We take, therefore, the same approach as \citet{fil88} and 
attempt to fit the Brackett lines with an accretion disk line profile.
Our line model is essentially the same as that discussed by
\citet{sto81} and \citet{sma81}.
The model disk is thin, circular, and in Keplerian motion around the
compact star.
We assume the line emission comes from an optically and physically
thin layer sitting on top of an optically thick disk.
The distribution of line emission across the disk is axially symmetric
and its radial dependence is given by a power law 
$F \propto r^{-\alpha}$.
The local line emission is Doppler broadened by isotropic
``turbulent'' velocities, which we describe by a Gaussian
distribution with a standard deviation
$\sigma = V_{turb}/V_{circ}$, where $V_{circ}$ is the local
circular velocity.
Finally, the emitting layer has a ratio of outer radius to inner radius
$R_{max}/R_{min}$.
These are the inner and outer boundaries of the region producing the
line emission, not necessarily the physical boundaries of the disk.
Thus the model line profile has three parameters:
$\alpha$,  $V_{turb}/V_{circ}$, and $R_{max}/R_{min}$.

\begin{figure}
   \figurenum{6}
   \includegraphics[angle=-90.0,scale=0.32]{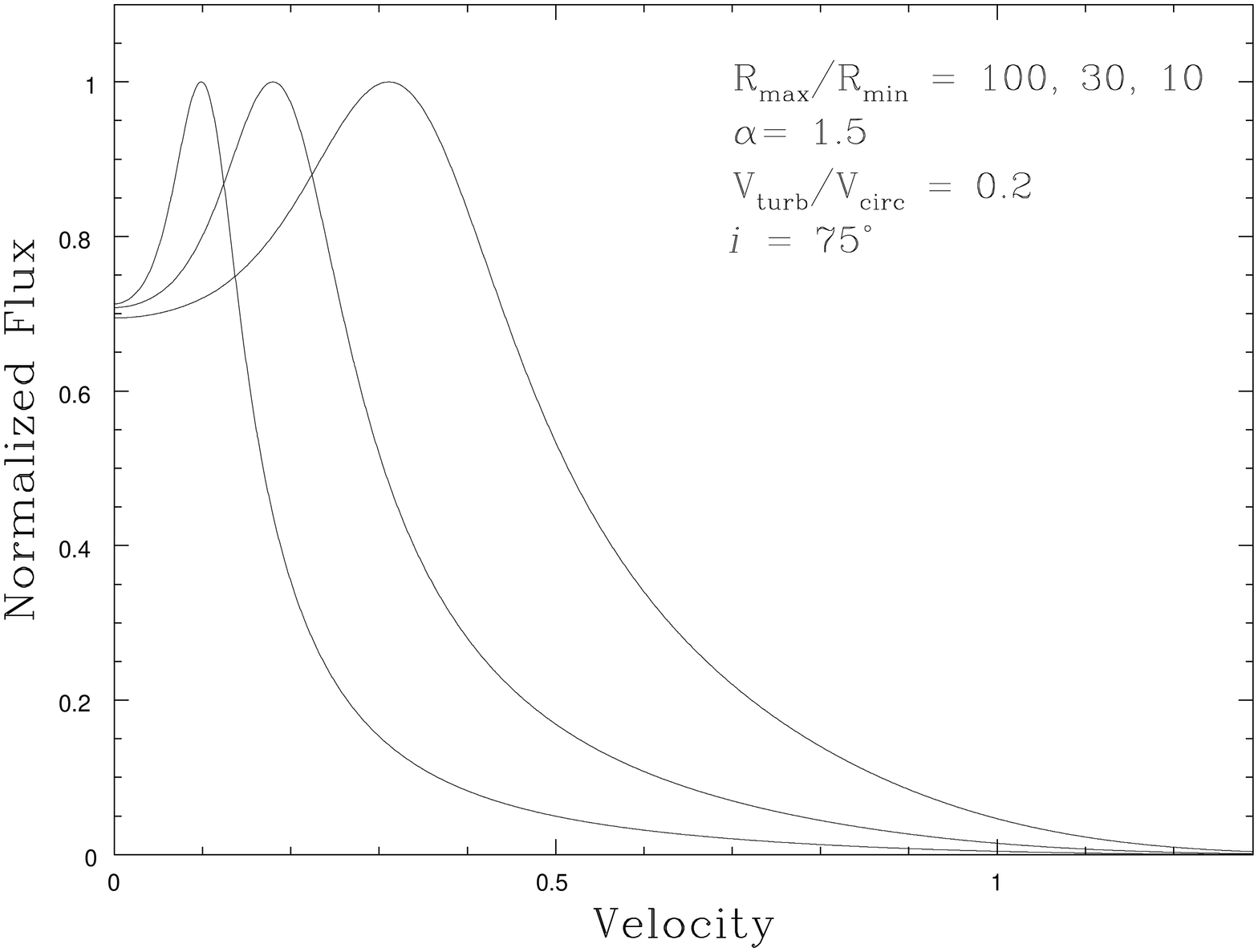}
   \includegraphics[angle=-90.0,scale=0.32]{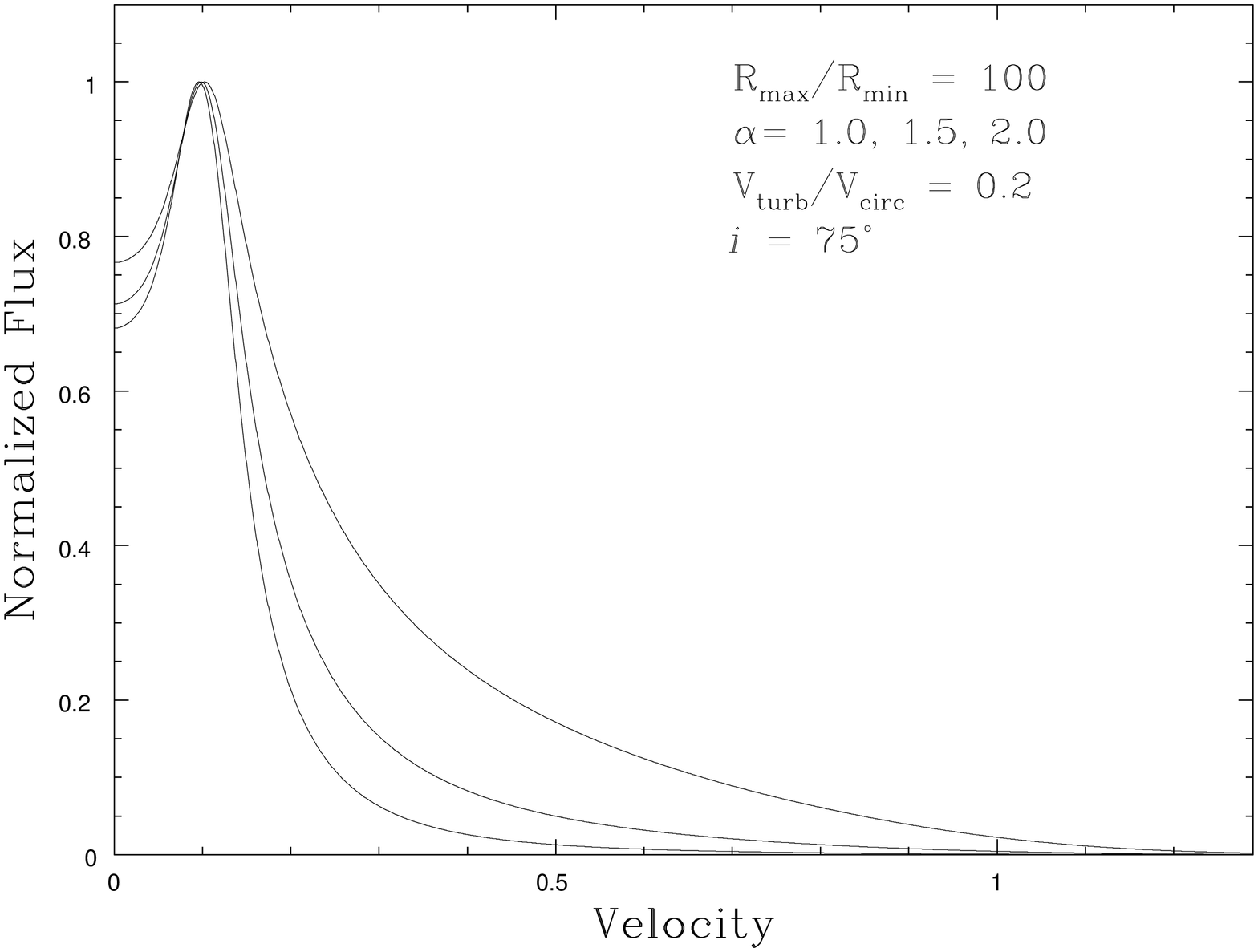}
   \newline
   \hspace*{1.5in}
   \includegraphics[angle=-90.0,scale=0.32]{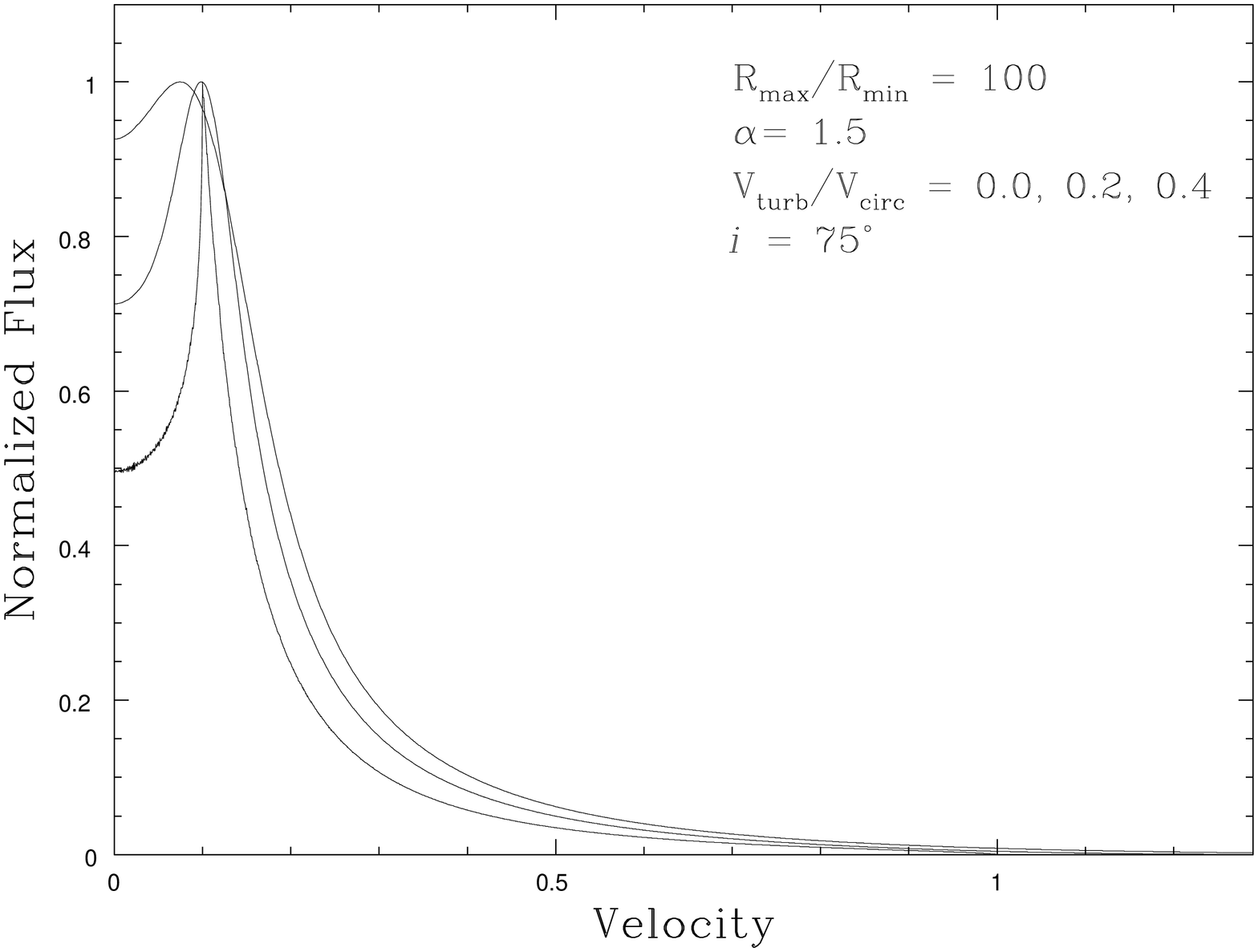}
   \caption{Model emission line profiles.
      The basic morphology of the profile -- 
      double-peaked with broad wings -- is independent
      of the model parameters over their ranges of interest.
      The upper left panel shows the effect of varying the parameter 
      $R_{max}/R_{min}$.
      The dominant effect is to increase the
      separation of the peaks relative to the total line width as
      $R_{max}/R_{min}$ decreases.
      The upper right panel shows the effect of varying the parameter
      $\alpha$, where the line flux from the disk is given by the power
      law $F \propto r^{-\alpha}$.
      The dominant effect is to throw more flux into the line wings as $\alpha$
      decreases.
      The lower panel shows the effect of varying the turbulent velocity
      of the gas in the disk.
      The dominant effect of changing $V_{turb}/V_{circ}$ is to decrease the
      depth of the minimum between the two peaks as the turbulent velocity
      increases.
}
   \label{FitParams-fig}
\end{figure}

Figure~\ref{FitParams-fig} shows how the model line profile varies as these
three parameters are varied.
The basic morphology of the profile -- 
double-peaked with broad wings -- is independent
of the parameters over their ranges of interest.
The dominant effect of varying $R_{max}/R_{min}$ is to change the
separation of the peaks relative to the total line width.
The dominant effect of changing $V_{turb}/V_{circ}$ is to change the
depth of the minimum between the two peaks.
The dominant effect of changing $\alpha$ is to change the 
relative amount of flux in the wings.
Another four parameters are needed to fit the model line profiles to 
the observed line:
the continuum level, the line strength, the wavelength of the 
line center, and a scale parameter for the line width that also absorbs
the effect of disk inclination.
Thus our model has a total of seven parameters (turbulent
broadening adds one parameter to the \citet{fil88} model).

The observed line profiles are affected by physical processes not
included in our model such as absorption, wind emission, and 
patchiness in the distribution of the emission whatever its source.
These effects systematically distort the observed line profiles,
vitiating any attempt to fit the profiles by, say, least squares.
Instead, we fitted the profiles by eye, emphasizing the
fits to the line wings.
This approach precluded a formal error analysis.

\subsection{Results}

\begin{figure}
   \figurenum{7}
   \hspace*{1.4in}
   \includegraphics[scale=0.45]{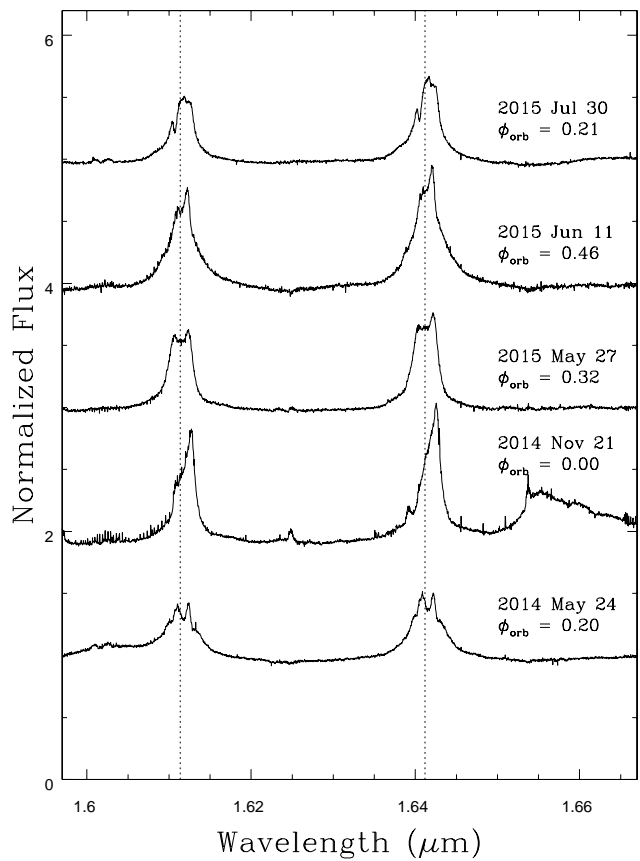}
   \caption{The Br12 and Br13 lines in all five spectrograms, labeled
      with the dates and the orbital phases at which the data were obtained.
      The forest of narrow features present in all the spectrograms, but 
      especially evident in the spectrogram obtained on 2014 November 21, 
      are caused by incompletely removed terrestrial absorption lines.
      Features near 1.625~$\mu$m are artifacts introduced by inaccurate
      flat-fielding at the ends of orders.
      The broad feature at 1.655~$\mu$m in the spectrum obtained on
      2014 November 21 is the jet line Br9$-$.}
   \label{Br12_13-fig}
\end{figure}

\begin{figure}
   \figurenum{8}
   \includegraphics[angle=-90.0,scale=0.32]{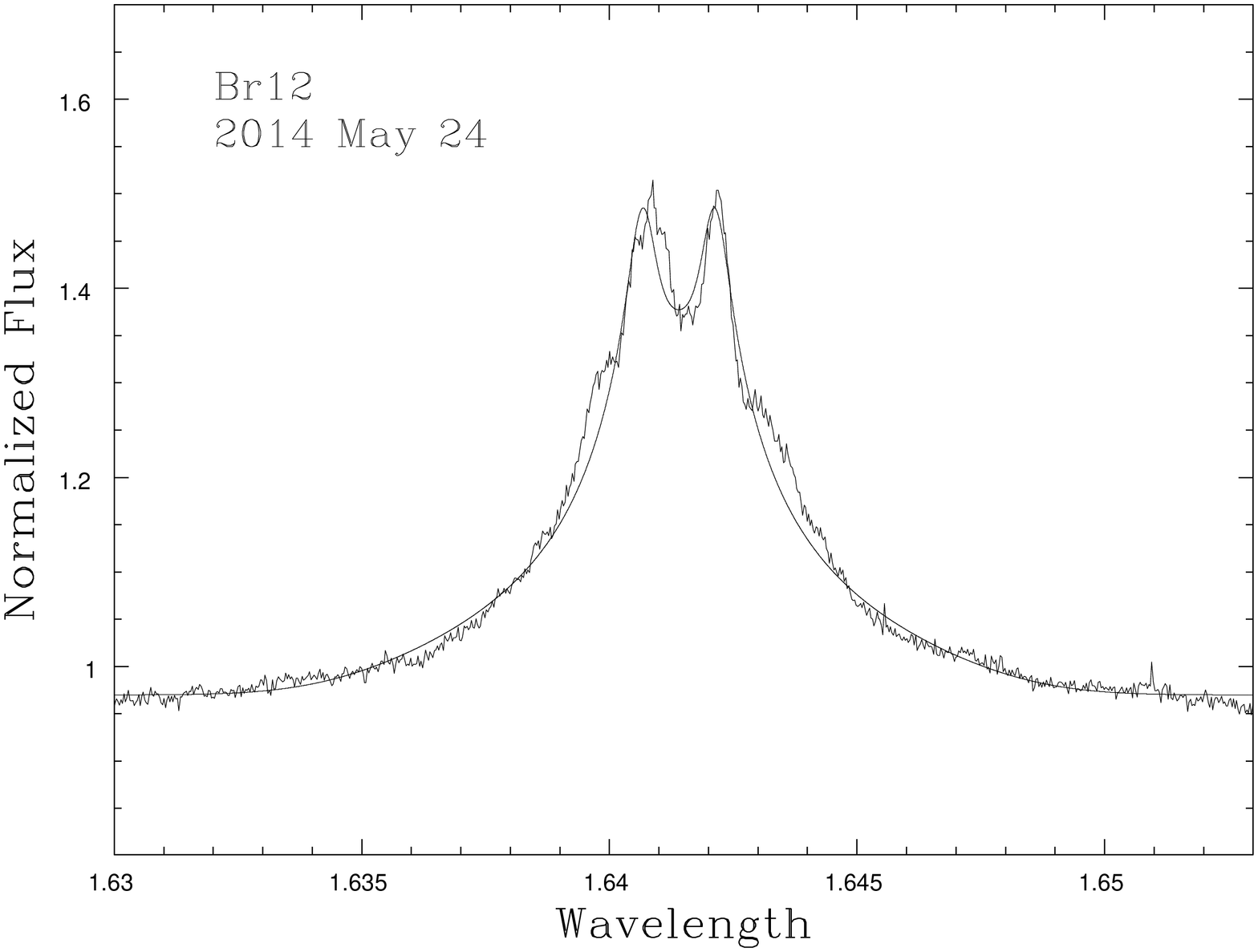}
   \includegraphics[angle=-90.0,scale=0.32]{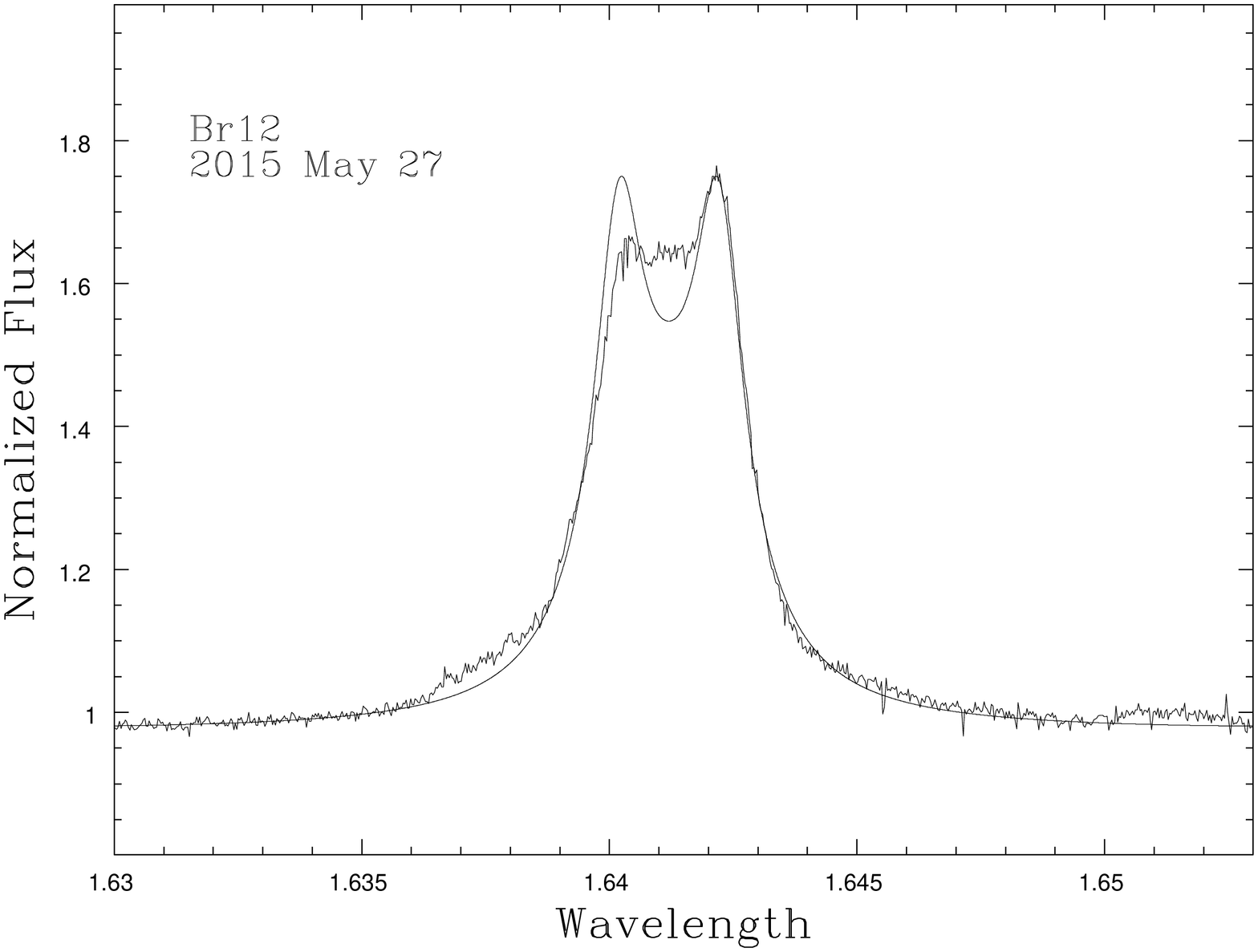}
   \newline
   \includegraphics[angle=-90.0,scale=0.32]{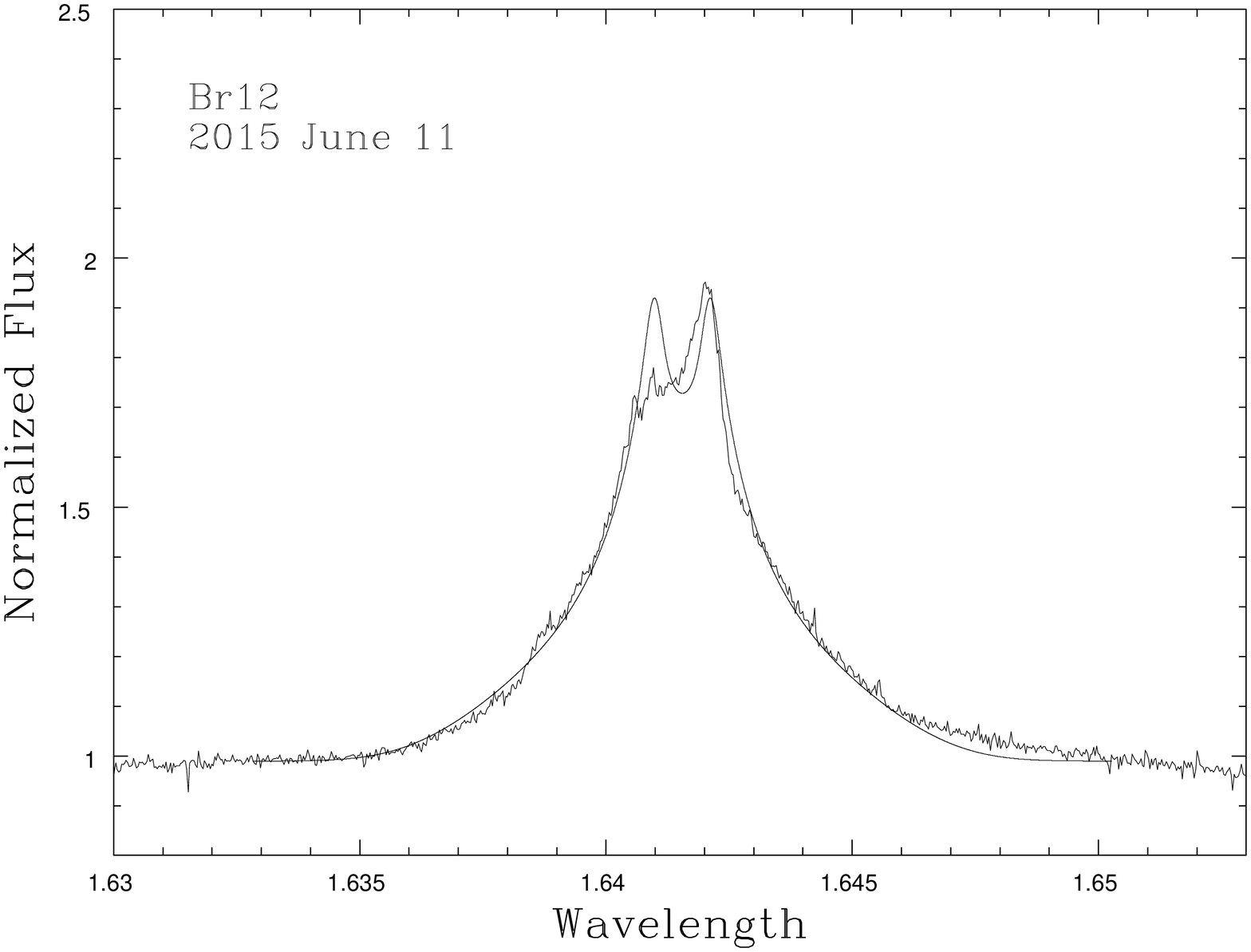}
   \includegraphics[angle=-90.0,scale=0.32]{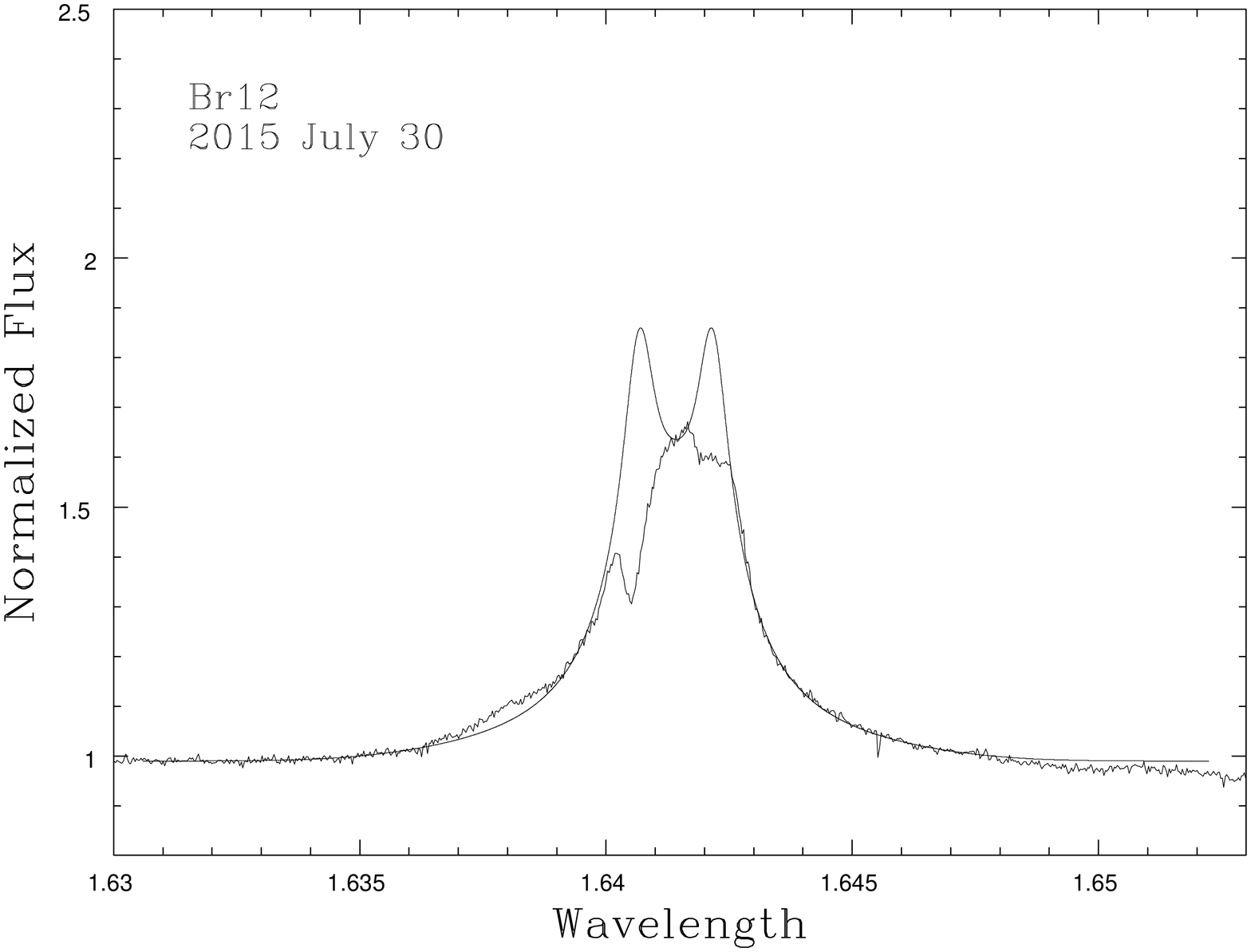}
   \caption{Fits of the accretion disk line profile model to the Br12 
      emission line on the dates
      labeled in the upper left corner of each panel.
      According to the \citet{eik01} ephemeris, the inclinations
      of the jet to the line of sight were 
      $\theta = 63^\circ \textrm{ and } 66^\circ$ in the upper two panels, and
      $\theta = 77^\circ \textrm{ and } 98^\circ$ in the lower two
      panels.
      If the accretion disk is perpendicular to the jet, the disk was viewed
      nearly edge-on in the lower right panel, perhaps explaining
      the missing horns of the line profile at that time.}
   \label{FourFits-fig}
\end{figure}

\begin{figure}
   \figurenum{9}
   \hspace*{1.5in}
   \includegraphics[angle=-90.0,scale=0.33]{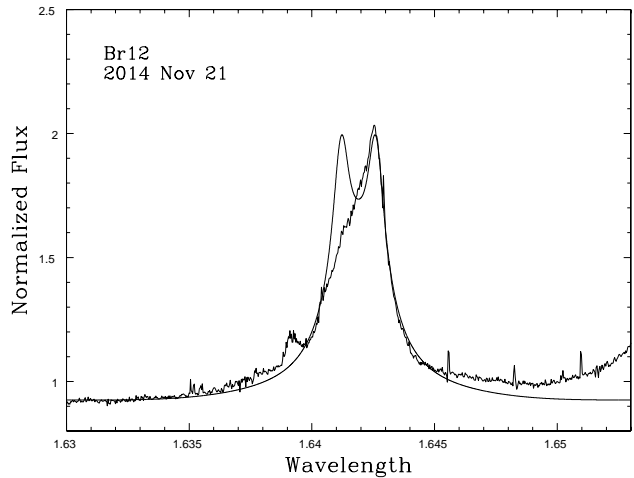}
   \caption{A fit of the accretion disk line profile model to the Br12 
      emission line on 2015 November 21.
      The rising flux on the right side of the spectrum is caused by the 
      jet line Br9$-$.
      According to the \citet{gor11} ephemeris, the orbital phase of SS 433 
      was $\phi_{orb}=0.00$ when the spectrogram was obtained.
      The profile of the Br12 line is similar to the profile of the Br$\gamma$
      line at $\phi_{orb}=0.96$ shown Figure~1 in \citet{per09}.
      Although the sample is small, we suggest that the distortion of the 
      line profile near and just before eclipse is repeatable.
      We interpret the distortion as being caused by a partial eclipse 
      of the accretion disk.}
   \label{FiveFit-fig}
\end{figure}

In part because Br$\gamma$ is contaminated by jet emission or
noise in all but two of our observations, but also because we
suspect the higher member of the Brackett series are less contaminated
by the wind emission seen in Br$\gamma$ by \citet{per09} and are more likely
to give a cleaner view of the accretion disk, we modeled only the 
higher members of the series.
Figure~\ref{Br12_13-fig} shows Br12 and Br13 from all our observations.
The Br12 and Br13 profiles agree well with each other, so we 
fit just the Br12 line, taking it as a proxy 
for all the higher members of the Brackett series.
The fits of the model profiles to the Br12 line are shown in 
Figures~\ref{FourFits-fig} and \ref{FiveFit-fig}, and the fitted
model parameters are listed in Table~\ref{ModelParams-tab}.
The table also gives the disk inclination, where we assume the disk
is perpendicular to the jet and have calculated
the jet inclination from the \citet{eik01} jet model.

We do not attach much significance to the fitted values of 
$R_{max}/R_{min}$ and $V_{turb}/V_{circ}$.
While the fitted values of $R_{max}/R_{min}$ are all near
$R_{max}/R_{min} = 100$, the values depend sensitively on the
placement of the continuum, on noise, and on systematic distortions
in the extreme wings of the line.
The large values of the ratio mean only that the emission
comes from most of the disk, not a narrow ring in the disk.
The values of $V_{turb}/V_{circ}$ all cluster tightly near 
$V_{turb}/V_{circ} = 0.20$, but
these values should also be viewed with skepticism.
The only
mechanism available to the model for filling the dip between
the peaks of the line is turbulent velocity.
Many other physical processes could do the same.
The rather high value of $V_{turb}/V_{circ}$ means only that
some mechanism is producing low velocity emission that tends
to fill the dip.

The fitted values of $\alpha$, which are dominated by the fits to the wings
of the lines, are more meaningful.
Except for the profile on 2014 November 21 (Figure~~\ref{FiveFit-fig}),
which is contaminated by a jet line, the disk model fits the wings of the 
observed profiles well.
The fitted values of $\alpha$ differ from observation to observation and 
these differences correspond to real differences among the observed profiles.
This can be seen by comparing the rapidly dropping wings of the profile on 
2015 May 27, which yielded $\alpha = 1.4$, to the slowly dropping 
wings of the profile on 2014 May 24, which yielded $\alpha = 2.1$.
All the values lie in the relatively narrow range $1.4 \leq \alpha \leq 2.2$,
bracketing the best-fit value $\alpha = 1.5$ found by
\citet{fil88}.
This range agrees with the range of values for $\alpha$ that have
been measured from the hydrogen emission lines produced by the accretion disks in
cataclysmic variables and low-mass X-ray binaries
\citep{sto81,sma81,joh89,hor91}.
Like SS~433, a single cataclysmic variable can also have different values of
$\alpha$ on different nights.
Thus, the value of $\alpha$ ranges from 1.7 to 2.25 in the dwarf nova Z~Cha
\citep{hor91}.

Although the disk model predicts that double peaks near the line center
should always be present, the peaks can be distorted or missing altogether
from the observed profiles of the Brackett lines.
Both peaks are missing from the Br12 profile on 2015 July 30 
(Figure~~\ref{FourFits-fig}).
We attribute this to absorption by the accretion disk. 
On that date the inclination of the disk was $98^\circ$, so the disk
was nearly edge-on to the Earth.
If, as is widely suspected, the disk is vertically extended \citep{fab04},
the rim of the disk would hide much of the emitting surface layer of the disk.
We expect our disk model to fail under these conditions, especially at the
low velocities where the hydrogen lines would be prone to self absorption.

The profiles of the Brackett lines on 2014 November 21 
(Figure~\ref{FiveFit-fig}) were single peaked --
the blue peak of the line was missing.
The orbital phase predicted by the \citet{gor11} ephemeris for
this observation is $0.004 \pm 0.004$.
\citet{per09} have published a montage of Br$\gamma$ line profiles.
One of them was obtained at a similar orbital phase ($\phi_{orb}=0.96$)
and was almost identical to the profile we observed.
Following the discussion in Section 2, the orbital phase of our observation
probably corresponds to a phase $-0.023$ before true conjunction of the stars.
Thus, our observation and the observation by \citet{per09} were both made 
during eclipse but prior to mid-eclipse.
At this phase the accretion disk is partly eclipsed and the
eclipse would systematically block off parts of the disk with
low velocity gas approaching the Earth.
The profile would be missing the blue-shifted peak.
The line profile can therefore be interpreted as a classical rotational disturbance 
similar to the rotational disturbance observed in the 
He~II~$\lambda$4686~\AA\ line from the accretion disk in the old nova
DQ~Her \citep{gre59,kra59}.

The separation of the peaks is approximately equal to twice
the projected orbital velocity at the outer edge of the disk.
We define the deprojected velocity at the outer edge
to be 1/2 the velocity separation of the peaks divided by the sine
of the disk inclination.
The deprojected velocities are listed in the penultimate column of 
Table~\ref{ModelParams-tab} and have a mean value of 148~km~s$^{-1}$.
The velocities on the individual observations range from
107 to 191~km~s$^{-1}$, and inspection of Figure~\ref{FourFits-fig}
shows that these differences correspond to large, real differences
in the peak separations from profile to profile.
In fact, these deprojected velocities may be too low because broadening 
of the lines by turbulence moves the double peaks towards
each other with respect to the line wings.
For $V_{turb}/V_{circ} = 0.20$ this effect reduces the separation of
the peaks by about 15\%.
Correcting for this effect would increase the mean deprojected velocity
at the outer edge of the disk to $\sim 170$~km~s$^{-1}$.
If, on the other hand, the shallowness of the dip between the peaks
has nothing to do with turbulent broadening, no correction is needed.
The most we can say is that the mean circular velocity at the
outer edge of the disk lies somewhere in the range 
150-170~km~s$^{-1}$.

\begin{figure}
   \figurenum{10}
   \hspace*{0.75in}
   \includegraphics[angle=-90.0,scale=0.50]{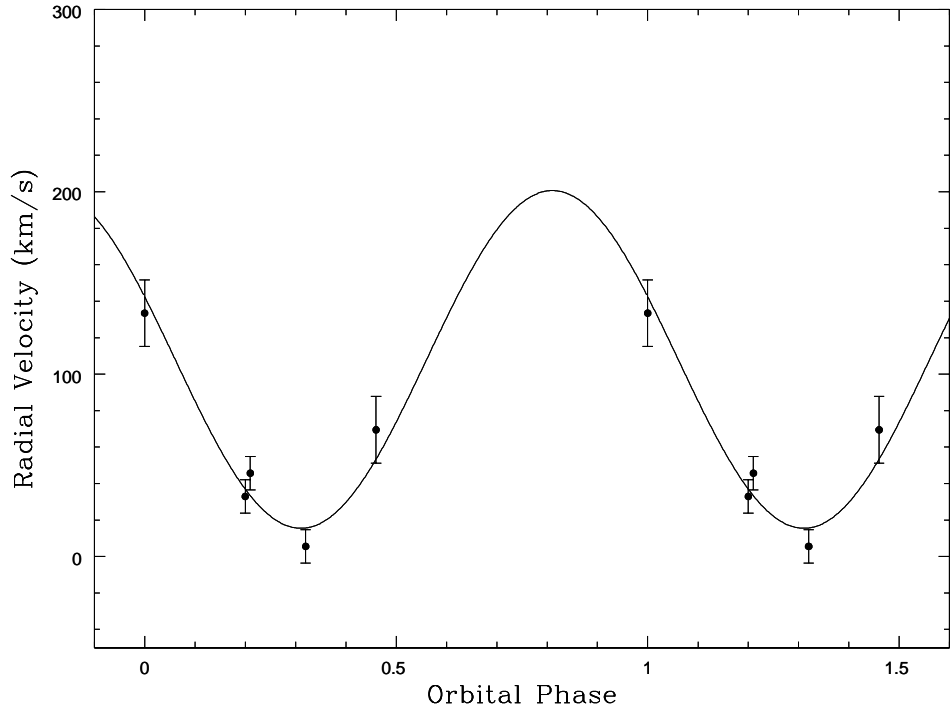}
   \caption{The data points are the radial velocities of the Br12 emission
      line plotted against the orbital phase calculated from the \citet{gor11}
      orbital ephemeris.
      If the Br12 emission line comes from a symmetric accretion disk around the
      compact star, its radial velocity should vary sinusoidally
      with a maximum approaching velocity at phase 0.25.
      The solid line is a sine curve fitted to the data points.
      The measured velocities are sparse and poorly distributed in orbital phase,
      so the values of the amplitude and mean velocity are not meaningful.
      The phase of the sine curve is, however, fairly well constrained and
      is $\phi_{orb} = 0.31 \pm 0.04$.}
   \label{RVcurve-fig}
\end{figure}

The wavelengths of the centers of the fitted profiles are given in 
the last column of Table~\ref{ModelParams-tab}.
Figure~\ref{RVcurve-fig} plots the radial velocities of the line
centers against the orbital phase calculated from the \citet{gor11}
orbital ephemeris (column 3 of Table~\ref{Obs-tab}).
If the Br12 emission line comes from an accretion disk around the
compact star, its radial velocity should be vary sinusoidally
with a maximum approaching velocity near phase 0.25.
The solid line in the figure is a sine curve fitted to the observed
velocities by least squares.
The measurements are sparse and poorly distributed in orbital phase,
so the fitted values of the amplitude and mean velocity of the sine
curve are not meaningful.
The phase of the sine curve is, however, fairly well constrained.
The phase of maximum approaching velocity occurs at
$\phi_{orb} = 0.31 \pm 0.04$, which is consistent with the value expected
for an accretion disk around the compact object.

\section{Summary and Discussion}

In summary, our most important results are:

\begin{itemize}
\item The precession phases of the jet differ from the phases predicted
by the \citet{eik01} ephemeris by $\Delta \Psi_3 \approx -0.03$.
Because this small phase shift could have been caused by a combination 
of jet nodding and intrinsic phase jitter, we conclude only 
that the \citet{eik01} ephemeris still predicts the jet redshift
phase to $|\Delta \Psi_3| \lesssim 0.03$.
This means that the clock underlying the 
phase of the jet precession has had a single, constant period 
for more than 37 years.
We can place an upper limit on any secular changes to the period
by attributing
all the observed phase shift to a slowly-changing precession period.
The implied upper limit to the rate of change of the
period is, then, $|\dot P_{jet}| \lesssim 2 |\Delta \Psi_3| / E^2
\approx 10^{-5}$, where $E = 83$ is the number of elapsed cycles, and
the time scale for any period change is
$P_{jet} /|\dot P_{jet}| \gtrsim 40,000$ years.
The superhumps in the light curves of the SU~UMa subclass of cataclysmic
variables are caused by precession of the accretion disks in these systems, 
although it is generally the precession of an elliptical disk lying flat in
the plane of the orbit, not a tilted disk \citep{pat01}.
For comparison, the rates of change of the periods of superhumps
are generally greater than $|\dot P_{hump}| \approx 2 \times 10^{-5}$,
but a few well-determined
values of $|\dot P_{hump}|$ are somewhat less than $10^{-5}$ \citep{kat09}.
The superhump periods of SU~UMa stars are typically just a few hours,
however, so for even the most stable of them
$P_{hump} / \dot P_{hump} \approx 30$ years.

\item The multicomponent, variable, P-Cygni profile of the 
He~I~$\lambda2.0587\ \mu$m emission line is direct evidence for an 
inhomogeneous and variable wind outflow.
The maximum observed wind speed is $\sim900$~km~s$^{-1}$.
These properties compare favorably with earlier measurements of the wind 
properties \citep{fab04,med13},
although we do not see the strong dependence of wind speed on disk
inclination found by \citet{fab97a}.
The He~II emission from the 12-7 and 10-7 transitions at $\lambda1.6898\ \mu$m 
and $\lambda2.1891\ \mu$m respectively also come from this wind.
We agree with \citet{med13} that the redshifted part of the He~II wind emission
is obscured, greatly obscured in the case of 
the $\lambda1.6898\ \mu$m line.
Because of this obscuration and because the obscuration is variable,
the profiles of the lines are distorted and the distortions
vary with time (see also 
the discussion of He~II emission lines in \citet{gor11}).

\item The broad, usually double-peaked, emission in the higher members
of the Brackett series are produced in
an accretion disk around the compact star.
Three lines of evidence lead to this conclusion.
First, the observed line profiles generally agree with model profiles  
produced by accretion disks.
We attribute the disagreements between the observed and model profiles
on two of the observing dates to a rotational disturbance during eclipse
and to obscuration by the optically-thick edge of the disk when 
the disk is edge-on, 
neither of which effects are included in the model.
Second, the derived distribution of line flux across the disk is 
similar to the distribution of line flux across the surfaces of disks 
in cataclysmic variables and low-mass X-ray binaries.
Third, the phase of the radial velocity curve of the Brackett lines
agrees with the phase expected for an accretion disk around the compact star.
Our conclusion is similar to that of \citet{fil88} who showed that the
higher members of the Paschen lines could be modeled by emission from 
an accretion disk.

Our model differs from essentially every published model for the Balmer
emission lines, especially the H$\alpha$ line \citep{fab04}.
We attribute the difference to a difference in the physical processes
producing the lines:
Emission from the wind dominates the lower members of the Balmer series,
drowning out emission from other sources.
Wind emission is much weaker in the higher members of the Paschen
series and Brackett series, allowing contributions from other components
of the SS~433 system to dominate.
Our model also disagrees with the interpretation of Br$\gamma$
by \citet{per09}.
In part this is due to our application of the model to the higher members
of the Brackett series, but another factor is also at work.
The observed Br12 line profile that differed the most from our model was obtained 
on 2015 July 30, when the accretion disk was almost edge-on and
the low-velocity line emission was strongly self absorbed.
All of the spectrograms obtained by \citet{per09} were obtained near
this same disk orientation.
They did not detect the accretion disk because much of it was hidden
at that precession phase.
\end{itemize}

Our results place an important limit the mass of the compact object.
We first review the uncertainties in previous measurements of
the masses of the stars in SS~433, then calculate the new upper
limit on the mass of the compact object.
The direct way to determine the masses of stars in binaries is to
measure the radial velocity curves of the two stars.
The most recent determination of the masses of the stars in
SS~433 based on 
radial velocities was by \citet{kub10}, who found
$1.9\, M_\odot \leq M_X \leq 4.9\, M_\odot$.
They also carefully analyzed the uncertainties in the masses derived
this way.
Because of heating,
the radial velocity of the A-star can be measured during
only half the orbit.
Furthermore, considerable judgement must be
exercised when choosing which of its absorption lines to 
measure and when estimating corrections to the measured velocities
to account for the heating.
The radial velocity of the compact star must be measured
from the emission lines arising in gas thought to follow 
its motion, usually He~II $\lambda 4686$~\AA\ 
or C~II $\lambda7231, 7236$ \AA\ \citep{fab90,gie02b}.
However, extensive observations of the radial velocity curves of soft X-ray
transients and catalclysmic variables have shown that 
emission-line radial velocities yield unreliable
masses, even when the systems are in unusually clean states,
such as dwarf novae in quiescence and soft X-ray transients at minimum light 
\citep{war95,mar98,cas06}.
As noted by \citet{gor11} and \citet{med13} and confirmed by
us, the He~II line arises mostly in the disk wind, not from
regions close to the compact star.
The profiles of the line are complex, highly variable, and have 
distortions correlated with orbital phase, introducing considerable
doubt that He~II is a good proxy for the motion of the compact star.

\citet{blu08} and \citet{per09} decomposed the H$\alpha$
and Br$\gamma$ emission lines into sets of up to six
Gaussian functions, and identified a pair of red- and 
blue-displaced Gaussians as emission from a circumbinary disk 
of gas in a circular Keplerian motion.
They equated the velocities of the Gaussians with the orbital
velocity, which yielded a lower limit on the mass enclosed by the 
circumbinary disk.
After subtracting the mass of the A-star, they derived a mass
of $\sim 16\, M_\odot$ for the compact star and its accretion
disk. 
Even if the decomposition of the line profiles into Gaussians were to
have physical meaning, the enhanced gas flow in the 
plane of the orbit is observed to have a large radial outflow velocity
\citep{cra81,fab97a,fab97b,fab04}.
The gas does not follow a circular Keplerian orbit, compromising masses
based on the assumption that it does.

Finally, as typified by the work of \citet{gor11,gor13}, 
it is possible to measure the mass of the compact
star in SS~433 without resorting to radial velocity measurements.
\citet{bri89} assumed that the A-star exactly fills its Roche lobe
and that the X-ray eclipse in SS~433 is an eclipse
by the A star of a thin jet of X-rays near the compact star.
Since the orbital inclination of SS~433 is known, they were
able to derive a mass ratio $q=M_X/M_A = 0.1496$, 
where $M_A$ is the mass of the A star.
Using only its photometric properties and distance,
\citet{gor13} derived a mass for the A-star and, from
the mass ratio, a mass for the compact star.
The assumption that the surface of A-star is identical to
the surface of its Roche lobe is insecure, though, especially since 
the A-star might be a tilted rotator as required by
slaved disk models for the jet precession \citep{van80}.
The derivation of the A-star mass from its photometric 
properties is, furthermore, a multi-step process, each 
step prone to its own uncertainties.

While we cannot fully determine the mass of the compact star in
SS~433, we can use the profiles of the Brackett lines to place an 
independent constraint on its mass.
The separation of the two stars in SS~433 is given by
$a^3 = G(M_X+M_A)P_{orb}^2/4\pi^2$.
Previous measurements of the masses of the stars tend to fall into two
distinct groups: one with total masses near $30\, M_\odot$ and one
with total masses near $15\, M_\odot$ \citep{kub10,gor11}.
For the $30\, M_\odot$ group, the separation of the two stars
is $\sim 5.1 \times 10^{12}$~cm;
while for the $15\, M_\odot$ group, the separation is 
$\sim 4.0 \times 10^{12}$~cm.
If we assume that the tidal truncation radius of an accretion disk 
around the compact object is about 25\% of the separation of the stars
\citep{fra92}, the maximum disk radius is $1.0 - 1.3 \times 10^{12}$~cm.

If the disk is Keplerian, the radius of the outer edge of the disk 
is given by $R_{max} = GM_X/V_{R_{max}}^2$, which leads to the mass limit
$M_X \leq R_{max} V_{R_{max}}^2 / G$.
All the various values one might adopt for $R_{max}$ and $V_{R_{max}}$
yield low upper limits to $M_X$, indicating a low total mass for SS~433.
We will, therefore, restrict our discussion to the smaller radius
for the disk, $R_{max} = 1.0 \times 10^{12}$~cm.
The most stringent constraint on the mass of the compact star comes
from adopting the smallest measured value of the deprojected rotational velocity, 
$V_{R_{max}}=107$~km~s$^{-1}$ as measured on on 2015 June 11.
This yields a mass limit of $M_X \leq 0.86\, M_\odot$.
A more realistic upper limit comes from adopting the mean deprojected
rotational velocity, $V_{R_{max}}=148$~km~s$^{-1}$,
which yields  $M_X \leq 1.6\, M_\odot$.
The most relaxed upper limit comes from increasing the adopted
$V_{R_{max}}$ by another 15\%, to $V_{R_{max}}=170$~km~s$^{-1}$, to
account for the possible effects of turbulent broadening on
the line profile.
In this case the mass limit increases to $M_X \leq 2.2\, M_\odot$.
The only recent mass determinations that are consistent with 
these limits are that of \citet{gor11}, who found 
$M_X = 1.45 \pm 0.20\, M_\odot$, and (barely) that of
\citet{kub10}, who found
$1.9\, M_\odot \leq M_X \leq 4.9\, M_\odot$.
These limits place the mass of the compact star
much below the range of measured black hole masses but
within the range of measured neutron star masses \citep{oze12}.

In closing we note that our results are based on data from just five nights
of observations.
It is encouraging that the new data
are in accord with previously published data where comparison is
possible, notably with the Paschen line profiles observed by
\citet{fil88} and with the Br$\gamma$ line profiles observed 
by \citet{per09}, at least at those orbital phases where our data overlaps
theirs.
Nevertheless, the observational properties of SS 433 are complex
and not easily disentangled.
The mass limit we have derive for the compact star
in SS~433 should be viewed with some caution until it is tested
against a larger set of data.

\acknowledgments 

This work used the Immersion Grating Infrared Spectrograph (IGRINS) that was
developed under a collaboration between the University of Texas at Austin and the Korea
Astronomy and Space Science Institute (KASI) with the financial support of the US
National Science Foundation under grant AST-1229522, of the University of Texas at
Austin, and of the Korean GMT Project of KASI.

\facility{McDonald Observatory (IGRINS).}


\clearpage
\begin{deluxetable}{cccc}
\tablecaption{Times and Phases of the Observations\label{Obs-tab}}
\tablewidth{6.5in}
\tablecolumns{4}
\tablehead{
   \colhead{Date (UT)} & 
   \colhead{Date (JD)\tablenotemark{a}} & 
   \colhead{$\phi_{orb}$\tablenotemark{b}}  &
 \colhead{$\Psi_3$\tablenotemark{c}} 
}
\startdata
2014 May 24 & 2456802.92728 & 0.198  & 0.88 \\
2014 Nov 21 & 2456983.55008 & 0.004  & 0.99 \\
2015 May 27 & 2457170.89553 & 0.325  & 0.15 \\
2015 Jun 11 & 2457185.79650 & 0.464  & 0.24 \\
2015 Jul 30 & 2457234.74350 & 0.206  & 0.54  
\enddata
\vspace{\baselineskip}
\tablenotetext{a}{Julian Day at the midpoint of the exposure sequence.}
\tablenotetext{b}{Orbital phase calculated from the ephemeris of \citet{gor11}.}
\tablenotetext{c}{Jet precession phase calculated from the ephemeris of Eikenberry et al. (2001)
but with the zero time translated so that $\Psi_3=0$ corresponds to the maximum blueshift
of the $z_{-}$ jet.  See Section 2 in the text.}
\end{deluxetable}

\clearpage
\begin{deluxetable}{ccccccc}
\tablecaption{Emission Lines in the Spectrum of SS 433 in the $H$ and $K$ Bands\label{LineIDs-tab}}
\tablewidth{7.5in}
\tablecolumns{6}
\tablehead{
\multicolumn{3}{c}{Stationary Lines} & & \multicolumn{3}{c}{Jet Lines}\\
   \cline{1-3}  \cline{5-7} \\
   \colhead{Species}     & 
   \colhead{Transition}  &
   \colhead{Wavelength}  &
                         &
   \colhead{Species}     & 
   \colhead{Transition}  &
   \colhead{Wavelength} \\
    & & ($\mu$m) &  & & & ($\mu$m)
}
\startdata
  Br$\gamma$    &      4-7                    &  2.1661          &  & P$\alpha$  &  3-4                       &  1.8756  \\
  Br8           &      4-8                    &  1.9451          &  & P$\beta$   &  3-5                       &  1.2822  \\
  Br10          &     4-10                    &  1.7367          &  &            &                            &          \\
 $\vdots$       &    $\vdots$                 &  $\vdots$        &  & Br$\beta$  &  4-6                       &  2.6259  \\ 
  Br25          &      4-25                   &  1.4971          &  & $\vdots$   &  $\vdots$                  &          \\
                &                             &                  &  & Br11       &  4-11                      &  1.6811  \\
  Pf19          &      5-19                   &  2.4490          &  &            &                            &          \\
  $\vdots$      &   $\vdots$                  &   $\vdots$       &  & He I       &  $3p ^3\! P^0 - 4d ^3\! D$ &  1.7007  \\
  Pf28          &      5-28                   &  2.3545          &  & He I       &  $3d ^3\! D - 4f ^3\! F^0$ &  1.8691  \\
                &                             &                  &  & He I       &  $3d ^3\! D - 4p ^3\! P^0$ &  1.9548  \\
   He I         &  $3p ^3\! P^0 - 4d ^3\! D$  &   1.7007         &  & He I       &  $3p ^3\! P^0 - 4s ^3\! S$ &  2.1126  \\
   He I         &  $3p ^3\! P^0 - 4s ^3\! S$  &   2.1126         &  &            &                            &          \\
   He I         &  $2s ^1\! S - 2p ^1\! P^0$  &   2.0587         &  & He I       &  $3d ^1\! D - 4p ^1\! P^0$ &  1.8556  \\
                &                             &                  &  & He I       &  $3p ^1\! P^0 - 4d ^1\! D$ &  1.9089  \\
  He II         &   12-7                      &   1.6923         &  & He I       &  $2s ^1\! S - 2p ^1\! P^0$ &  2.0587  \\
  He II         &   10-7                      &   2.1891         &  &            &                            &       \\
                &                             &                  &  &            &                            &       \\
  Mg II         & $5s ^2\! S_{1/2} - 5p ^2P^0_{3/2}$ & 2.1375    &  &            &                            &       \\
  Mg II         & $5s ^2\! S_{1/2} - 5p ^2P^0_{1/2}$ & 2.1438    &  &            &                            &       \\
\enddata
\end{deluxetable}

\clearpage
\begin{deluxetable}{@{\enspace}@{\extracolsep{15pt}}cccc}
\tablecaption{Observed Redshifts of the Jet Emission Lines\label{RedShifts-tab}}
\tablewidth{5.0in}
\tablecolumns{4}
\tablehead{
 \colhead{Date (UT)} & 
 \colhead{$\Psi_3$} & 
 \colhead{$\Delta z_{+}$} & 
 \colhead{$\Delta z_{-}$}
}
\startdata
2014 May 24  & 0.88 & 0.1510  & -0.0775 \\
2014 Nov 21  & 0.99 & 0.1640  & -0.0882 \\
2015 May 27  & 0.15 & 0.1200  & -0.0465 \\
2015 Jun 11  & 0.24 & 0.0699  & \nodata \\
2015 Jul 30  & 0.54 & -0.0090 & 0.0760 
\enddata 
\end{deluxetable}

\clearpage
\begin{deluxetable}{ccccccc}
\tablecaption{Fitted Parameters for the Disk Emission Line Model\label{ModelParams-tab}}
\tablewidth{5.0in}
\tablecolumns{7}
\tablehead{
    Date       &  & & & Disk & $V_{R_{max}}$\tablenotemark{b} & Line Center \\
 \colhead{ (UT)                     } & 
 \colhead{ $R_{max}/R_{min}$             } & 
 \colhead{ $\alpha$                      } & 
 \colhead{ $V_{turb}/V_{circ}$           } &
 \colhead{ Inclination\tablenotemark{a}  } &
 \colhead{ (km~s$^{-1}$)                 } &
 \colhead{ ($\mu$m)                      }
}
\startdata
2014 May 24  &  100  &  2.1  &  0.20  & $63^\circ$ & 147     & 1.64135(05 )\\
2014 Nov 21  &  110  &  1.7  &  0.20  & $57^\circ$ & 148     & 1.64190(10) \\
2015 May 27  &  100  &  1.4  &  0.22  & $66^\circ$ & 191     & 1.64120(05) \\
2015 Jun 11  &  100  &  2.2  &  0.20  & $77^\circ$ & 107     & 1.64155(10) \\
2015 Jul 30  &\ \ 80 &  1.7  &  0.22  & $98^\circ$ & \nodata & 1.64142(05)
\enddata 
\vspace{\baselineskip}
\tablenotetext{a}{The disk is assumed to be perpendicular to the
 jet, whose inclination is determined from the jet precession ephemeris.
 At an inclination of $90^\circ$ the disk is edge-on.}
\tablenotetext{b}{Deprojected orbital velocity at the outer edge of the disk.}
\end{deluxetable}

\clearpage

\clearpage

\clearpage

\clearpage

\clearpage

\clearpage

\clearpage

\clearpage

\clearpage

\clearpage

\end{document}